\documentclass[useAMS,usenatbib,usegraphicx]{mn2e}
\usepackage{amssymb}
\usepackage{amsmath}
 \usepackage{color}
\usepackage{soul}

\DeclareMathOperator{\rot}{rot}
\DeclareMathOperator{\Div}{div}

\title{The spin evolution of neutron stars with the superfluid core.}
\author[D. P. Barsukov, O. A. Goglichidze and A. I. Tsygan]{D. P Barsukov\thanks{E-mail:
bars@astro.ioffe.ru}$^{1,2}$, O. A. Goglichidze\thanks{E-mail:
goglichidze@gmail.com}$^1$ and A. I. Tsygan\thanks{E-mail:
tsygan@astro.ioffe.ru}$^1$\\
      $^1$Ioffe Physical-Technical Institute of the Russian Academy of Sciences, Saint-Petersburg, Russian Federation \\
      $^2$Saint-Petersburg State Polytechnical University, Saint-Petersburg, Russian Federation}
\begin{document}
  \maketitle
  \begin{abstract}
    We investigate the neutron stars spin evolution (breaking, inclination angle evolution and radiative precession), taking into account the superfluidity of the neutrons in the star core. The neutron star is treated as a two-component system consisting of a ``charged'' component (including the crust and the core protons, electrons and normal neutrons) and a core superfluid neutron component. The components are supposed to interact through the mutual friction force. We assume that the ``charged'' component rotates rigidly. The neutron superfluid velocity field is calculated directly from linearized hydrodynamical equations. It is shown that the superfluid core accelerates the evolution of inclinaton angle and makes all pulsars evolve to either orthogonal or coaxial state. However, rapid evolution seems to contradict the observation data. Obtained results together with the observations may allow to examine the superfluid models.
  \end{abstract}
  \begin{keywords}
    dense matter -- hydrodynamics -- stars:neutron -- pulsars:general
  \end{keywords}
  
  \section{Introduction}
%   Neutron stars are the highly magnetised, rapidly rotating compact astrophysical objects. Most of them have been detected as a radio pulsars, the sources of the pulsing radio signals, the nature of which is explained with the lighthouse model. The observed radiation emits in the narrow cones from the open field lines regions of the pulsar magnetosphere located above the magnetic poles, and the radiation can only be observed when the beam of emission is pointing towards the Earth.  
  To date there is no theory explaining the pulsars emission mechanism but it is clear that the basic parameters of pulsars signals such as the width of the pulses profiles depend on the spatial orientation of at least two vectors: the angular velocity vector $\vec{\Omega}$ and the magnetic moment vector $\vec{m}$ which determines the emission direction.
  During the neutron star life the magnitudes as well as the relative orientation of this vectors can evolve. 
  
  Pulsars rotation dynamics is a complex process. As it was pointed by \citet{GoldreichJulian1969}, neutron stars must possess a very large magnetosphere. Strictly speaking, the torque acting on the neutron star can be calculated only by using the self-consistent theory of the magnetosphere which is far from complete at present \citep{Spitkovsky2008}. It makes researchers use the various model expressions for the neutron star angular momentum loses torque. The simplest one is the 
  radiation-reaction torque, acting on the magnetised sphere which rotates in a vacuum \citep{Deutsch1955}. In that case, the combination $P/\cos\chi$, where $P$ is the pulsar rotation period and $\chi$ is the inclination angle (the angle between the spin axis and the magnetic moment), remains constant during the pulsar lifetime \citep{DavisGoldstein1970,MichelGoldwire1970}. It leads to a very fast magnetic axis alignment, which appears to contradict the observations. This result holds for a fluid star whose 
  rotation deformation follows the instantaneous rotation axis.  
  
  However, this contradiction seems to be a result of the excessive idealization.  
  If the neutron star shape is not perfectly spherical and it is not symmetrical about the magnetic axis, the precession caused by star oblateness will prevent the pulsar alignment \citep{Goldreich1970}.
  
  Except the magnetic-dipolar angular momentum loses mechanism pulsars have another one
  related with the charge particles currents in the open field lines regions (pulsar tubes) of the neutron star magnetosphere \citep{Jones1976}. Currents mechanism forces the pulsar to counteralign, and if it is comparable with the magnetic-dipolar mechanism, owing to their competition the inclination angle evolves much more slowly than it follows from the vacuum approximation \citep{BGT2013}.  
  
  In the presence of a magnetosphere the radiation-reaction torque itself may, in principle, differ significantly from the vacuum case. For example, in the framework of the force-free approximation \citet{BeskinGurevichIstomin1983} have argued that if the tubes currents are absent, a pulsar does not lose its angular momentum through the magnetic-dipolar mechanism at all.
  
  Neutron stars are not rigid perfectly, so the different dissipation mechanisms may take place in their interiors. Because of the finite rigidity the star crust changes its shape during the precession period. It leads to the damping the precession motion and decreasing the precession angle \citep{ChauHenriksen1971,Macy1974}. 

  Neutron stars cores consist mostly of the neutrons with small fractions of protons, electrons and probably more exotic particles like hyperons, kaons, etc. The neutrons in some regions of the star should be in superfluid state. Superfluidity may appear in the rotational dynamics in several ways at several time-scales. First, superfluidity is believed to be responsible for pulsars glitches, the sudden changes of the observational pulsars periods with subsequent smooth recovery to the initial value. The recovery time-scales are in the range on days to  months.
  As it was proposed by \citet{BaymPethickPines1969b}, the central role here plays the pinning and sudden unpinning of the vortices governing the superfluid rotation to the neutron star crust nuclear lattice.

  The vortices pining may also cause a neutron star precession. \citet{Shaham1977} has shown that for the neutron stars with low oblateness the precession frequency is given by $\omega_p = \Psi \Omega$, where $\Psi$ is the ratio of moment of inertia of the pinned superfluid to the rigid component of the rest of the star. The such precession must be very fast. The observational values of the precession periods obtained for several pulsar, contrary to the theory, all are of the order of a year. The more detailed theories including the unperfect pinning do not make the situation more clear (see \citet{Link2006} for the detail).  
  
  Long-term rotation dynamics taking into account the internal structure of the neutron stars was studied by researchers too.
  Crust-core coupling has been investigated by \citet{Easson1979}, who has shown that the crust and the core plasma are very close to co-rotation which can be supported either by the internal magnetic field or by the viscosity. It should be noted, however, that he neglected the interaction between the core proton-electron plasma and the superfluid neutrons. 
  
  The influence of core neutrons superfluidity on the spin evolution was studied by \citet{CasiniMontemayor1998}. They treat a neutron star as a two component system the components of which rotate rigidly with angular velocities $\vec{\Omega}_c$ and $\vec{\Omega}_0$.  Crust component interacts with core by the friction-like torque $\vec{N} = -\tau_f^{-1}(\vec{\Omega}_c - \vec{\Omega}_0)$ and feels the action of an external torque. The similar problem was investigated by \citet{BGT2013} but they used the more correct expression for $\vec{N}$ obtained directly  by the integration of the mutual friction force \citep{SedrakianWassermanCordes1999}.

  The present paper considers the development of the semi-hydrodynamical formalism (only superfluid neutrons differential rotation effects are taken into account) describing the spin evolution of neutron stars which allows to include in the equations the realistic mass densities and interaction coefficients profiles as well as the different external torques. This work extends  \citet{BGT2013} (and \citet{CasiniMontemayor1998}) and it is a step toward the full hydrodynamical formalism.
  
  The paper is organised as follows. In section \ref{sec:s_hydrodynamics} we review the basic features of the superfluid mixture hydrodynamics. The proposed semi-hydrodynamical model is described in section \ref{sec:model}. Here, we obtain the expression for non-axisymmetric core neutrons superfluid flow. The equations of motion (described the star breaking, inclination angle evolution and radiative precession) are formulated in section \ref{sec:eq_of_motion}. In section \ref{sec:application} we apply some particular models of the neutron superfluidity, neutron star core constituents interaction and the particular model of the external torque. Obtained results is discussed in section \ref{sec:discussion}.

  \section{Superfluid hydrodynamics}\label{sec:s_hydrodynamics}
  It is well known that for the uncharged superfluids the velocity field $\vec{v}$ must satisfy the equation $\rot \vec{v} = 0$. However, the superfluids can rotate by forming an array of vortices. The velocity field near the each vortex has the form \citep{TilleyTilley_book}:
  \begin{equation}
    \vec{v} = \frac{\hbar}{2 m_{N} \tilde{r}_v}\vec{e}_{v\phi}, 
  \end{equation}  
  where $m_{N}$ is the mass of nucleon forming the superfluid, $\tilde{r}_{v}$ is the distance from the vortex core, $\vec{e}_{v\phi}$ is the azimuthal unit vector. Vorticity is contained only in the vortex cores, in which the superfluidity breaks:
  \begin{equation}
    \label{eq:v_def}
    \rot\vec{v} = \sum\limits_i\int\varkappa\vec{e}_k\delta\left(\vec{r}_{v_i}(l)-\vec{r}\right)dl
  \end{equation}
  \citep{Sonin1987}. Here, $\vec{e}_k$ is the unit  vector pointing vortices orientation, $\vec{r}_{v_i}(l)$ is the $i$-th vortex line equation and $\delta(\vec{r})$ is the three-dimensional Dirac delta function. Each vortex carries the quantum of circulation $\varkappa = 2\pi\hbar/2m_N$.
  
  However, if the characteristic scales much larger than the intervortex space $l_v$ (long wavelength limit), an averaging procedure can be applied (see \citet{BaymChandler1983} 
  for the detail). One can introduce a  smooth vorticity field $\vec{\omega}_s = n_v \varkappa\vec{e}_k$, where $n_v$ is the number of vortices per unit area. The conservation law for $\vec{\omega}_s$ has the form
  \begin{equation}
    \label{eq:vorticity_conservation}
    \partial_t\vec{\omega}_s + \rot[\vec{\omega}_s \times \vec{v}_l] = 0.
  \end{equation}
  Here, $\vec{v}_l$ is the velocity of the vortices.
  One can formally introduce a vector field $\vec{v}_s$ such that 
  \begin{equation}
    \label{eq:OmsVs}
    \vec{\omega}_s = \rot\vec{v}_s. 
  \end{equation}
  Substituting this into eq. \eqref{eq:vorticity_conservation}, 
  it is easy to verify that the last is equivalent to the Euler-like equation
  \begin{equation}
    \label{eq:Euler_simple}
    \partial_t\vec{v}_s+(\vec{v}_s\cdot\nabla)\vec{v}_s+\nabla\tilde{\mu} = \vec{f}_v,
  \end{equation}
  where
  \begin{equation}
    \label{eq:force_simple}
    \vec{f}_v = -\vec{\omega}_s\times(\vec{v}_l-\vec{v}_s)   
  \end{equation}
  and $\tilde{\mu}$ is some scalar function. 
%   \textbf{This arbitrariness arises from the ambiguity of definition of $\vec{v}_s$ field.} 
  Note that $\vec{v}_s$ is given up to gradient of an arbitrary function. The replacement $\vec{v}_s \rightarrow \vec{v}_s+\nabla\vartheta$  leads to replacement $\tilde{\mu}\rightarrow\tilde{\mu}-\partial_t\vartheta-\vec{v}_s\cdot\nabla\vartheta-(\nabla\vartheta)^2/2$.
%   \textbf{The arbitrariness of $\vec{v}_s$ derived from \eqref{eq:OmsVs} results just in the redefinition of $\tilde{\mu}$.}
  Equation \eqref{eq:v_def} allows us to uniquely determine $\vec{v}_s$ as $<\vec{v}>$, where $< >$ denotes the averaging over unit area. In this case, function $\tilde{\mu}$ should be a chemical potential per unit mass \citep{BaymChandler1983}.
  
  In order to exclude $\vec{v}_l$ from the equation let us consider the force balance for a vortex. 
  On the one hand, if the superfluid coexists with some liquid which moves with velocity $\vec{v}_c$ (index ``c'' is used in order to unify the notations with the following sections of the paper), the vortex feels the action of the friction force. 
%   It can be the thermal excitation fluid or any liquid of another kind.
  A lot of friction mechanisms can take place in the neutron star interior but, in any case, if the friction is not very strong, the force acting per unit length can be represented as
  \begin{equation}
    \label{eq:fric_force}
    \vec{F}_c = -\frac{\rho_c}{n_v\tau_c}(\vec{v}_l-\vec{v}_c),
  \end{equation}
  where $\rho_c$ is density of ``c''-liquid, $\tau_c$ is the characteristic time-scale depending on particular friction mechanism. If there are several comparable mechanisms in some star region,  $\tau_c$ is calculated by the formula $\tau_c^{-1} = \tau_{c1}^{-1}+\tau_{c2}^{-1}+...+\tau_{ck}^{-1}$, where $\tau_{ci}$ are the time-scales specifying each mechanism separately.

  On the other hand, if the vortex moves through the superfluid, it feels the action of the Magnus force which has the form
  \begin{equation}
    \label{eq:Magnus_force}
    \vec{F}_m = \rho_s \varkappa\vec{e}_k \times(\vec{v}_l-\vec{v}-\rot\vec{\lambda}).
  \end{equation}
  Here, $\rot\vec{\lambda}$ is the term arising from the vortex local self-acting (see, for example, \citet{Schwarz1985}), $\vec{\lambda}$ is a vector parallel to $\vec{e_k}$.

  The thickness of the vortex core equals to the coherent length $\xi_n$ which for the neutron star core is of the order of several tens nucleon sizes \citep{BaymPethickPines1969a}. The mass contained in the vortex core are negligibly small. Thus, Newton second law for the vortex reduces just to
  \begin{equation}
    \label{eq:vortex_newton_law}
    \vec{F}_m+\vec{F}_c = 0.
  \end{equation}
  
  Solving equation \eqref{eq:vortex_newton_law} for $\vec{v}_l$, averaging the result over unit area and substituting it into \eqref{eq:force_simple}, one can obtain \citep{SedrakianWassermanCordes1999, SideryAlpar2009} 
  \begin{align}
    \nonumber
    \vec{f}_v = -\vec{\omega}_s\times\rot\lambda_n + \beta'\vec{\omega}_s\times(\vec{v}_s-\vec{v}_c+\rot\lambda_n)+ \\ +\beta\vec{e}_k\times\left[\vec{\omega}_s\times(\vec{v}_s-\vec{v}_c+\rot\lambda_n)\right],
  \end{align}
  where we introduce two coefficients
  \begin{equation}
    \label{eq:betas_def}
    \beta = \frac{\sigma}{1+\sigma^2}, \ \ \beta' = \frac{\sigma^2}{1+\sigma^2},
  \end{equation}  
  where 
  \begin{equation}
    \sigma = \frac{\rho_c}{\rho_s}\frac{1}{\tau_c\omega_s}
  \end{equation}
  is the coupling parameter.
  This kind of interaction taking place in superfluids is usually called the mutual friction \citep{HallVinen1956}. Depending on  $\sigma$ there are two regimes. Weak-coupling regime corresponds to $\sigma\ll1$. In this case, $\beta^2\approx\beta'\ll1$ and $\vec{v}_l\approx\vec{v}_s$. If the opposite inequality $\sigma\gg1$ satisfies, the strong-coupling regime takes place. It means that
  $\beta\ll1$, $\beta'\approx1$ and $\vec{v}_l\approx\vec{v}_c$. Note that this analysis is quite general. 
  The only thing which determines by the physical nature of ``c''-liquid is $\tau_c$.
%   The physical nature of ``c''-liquid specified [determined] here only by $\tau_c$. 

  So, in the long wavelength limit there is no need to consider the vortices dynamics. Instead that one can solve the Euler-like equation to which the mass, momentum and energy conservation laws for the whole liquid (including thermal excitations) should be added of course.

  The hydrodynamical equations for the rotational superfluid liquids can be obtained from the conservation laws consideration \citep{Khalatnikov_book}. This method don't deal with the vortex lattice at all, but it is quite rigorous and it naturally can be extended to the more general superfluid systems. One just need to assume that the internal energy depends on the $\vec{\omega}$.

  An example of such systems is a mixture of superfluid liquids. \citet{AndreevBashkin1976} have shown that the superfluid hydrodynamics must be modified in that case. The strong interaction between the particles forming the two superfluid constituents gives rise to the so-called entrainment effect which leads to the following relations: 
  \begin{align}
    \label{eq:neutron_current}
    &\vec{J}_n = (\rho_n-\rho_{nn}-\rho_{np})\vec{v}_{ex}+\rho_{nn}\vec{v}_{n}+\rho_{np}\vec{v}_{p}, \\
    \label{eq:proton_current}
    &\vec{J}_p = (\rho_p-\rho_{pp}-\rho_{pn})\vec{v}_{ex}+\rho_{pp}\vec{v}_{p}+\rho_{pn}\vec{v}_{n}, \\
    \label{eq:neutron_momentum}
    &\vec{p}_n = (\rho_n-\rho_{nn}-\rho_{np})\vec{v}_{ex}+(\rho_{nn}+\rho_{np})\vec{v}_{n}, \\
    \label{eq:proton_momentum}
    &\vec{p}_p = (\rho_p-\rho_{pp}-\rho_{pn})\vec{v}_{ex}+(\rho_{pp}+\rho_{pn})\vec{v}_{p},
  \end{align}
  where $\vec{J}_\alpha$ is the mass current of the $\alpha$ constituent ($\alpha = p,n$), $\vec{p}_\alpha$ are the momentum densities, $\vec{v}_{ex}$ is the thermal excitation velocity (it is assumed that the excitation velocity is the same for both fluids), $\rho_{\alpha\beta}$ is the mass density matrix (also called the entrainment matrix or the Andreev-Bashkin matrix). It can be shown that $\rho_{np} = \rho_{pn}$.  So, in the superfluid mixtures the mass currents corresponding to each fluid are no longer parallel to they velocities and don't equal to corresponding momentum density. Note, however, that 
  \begin{equation}
    \label{eq:J&p_eq}
    \vec{J}_n+\vec{J}_p = \vec{p}_n+\vec{p}_p.
  \end{equation}
  The superfluid velocities here are determined just like for the ordinary superfluids:
  \begin{equation}
    \vec{v}_n = \frac{\hbar}{2m_n}\nabla S_n, \ \ \vec{v}_p = \frac{\hbar}{2m_p}\nabla S_p - \frac{e}{m_p c}\vec{A},
  \end{equation}
  where $S_\alpha$ are the phases of the complex order parameters, $\vec{A}$ is the electromagnetic vector potential.

  One can also introduce the mass currents and momentum densities for the superfuids and thermal excitations separately:
  \begin{align}
    \label{eq:sf_neutron_current}
    &\vec{J}_s = \rho_{nn}\vec{v}_{n}+\rho_{np}\vec{v}_{p}, \\
    \label{eq:sf_proton_current}
    &\vec{J}_{p(s)} = \rho_{pp}\vec{v}_{p}+\rho_{pn}\vec{v}_{n}, \\
    \label{eq:sf_neutron_momentum}
    &\vec{p}_s = \rho_s\vec{v}_{n}, \\
    \label{eq:sf_proton_momentum}
    &\vec{p}_{p(s)} = \rho_{p(s)}\vec{v}_{p},\\
    &\vec{p}_{ex} = \rho_{ex}\vec{v}_{ex},
\end{align}
  where we denote $\rho_s = \rho_{nn}+\rho_{np}$, $\rho_{p(s)} = \rho_{pp}+\rho_{pn}$ and $\rho_{ex} = \rho_n+\rho_p-\rho_{s}-\rho_{p(s)}$. Here and after we will use the index ``$s$'' instead ``$n(s)$'' for all quantities relating to the superfluid part of the neutron liquid.

  The system of hydrodynamical equations describing the mixture of superfluid neutrons, superconducting protons and degenerate electrons and muons taking into account the neutron-proton entrainment and gravitational force was developed by Mendell and Lindblom \citep{MendellLindblom1991,Mendell1991a, Mendell1991b}. They used the Khalatnikov's method.
  The ``superfluid neutron'' part of Mendell-Lindblom system of equations has the form
  \begin{equation}
    \label{eq:Mendell_Euler}
    \partial_t\vec{v}_s+(\vec{v}_s\cdot\nabla)\vec{v}_s+\nabla(\tilde{\mu}+\Phi_G) = \frac{\rho_{np}}{\rho_{s}} (\vec{v}_p-\vec{v}_s)\times\vec{\omega}_s+\vec{f}_v,
  \end{equation}
  \begin{equation}
    \label{eq:Mendell_continuity}
    \partial_t\rho_s + \Div\vec{J}_s = \Gamma_s,
  \end{equation}
  \begin{align}
    \label{eq:Mendell_force}
     &\vec{f}_v=-\frac{1}{\rho_s}\vec{\omega_s}\times\rot\vec{\lambda}_n+\\
     \nonumber
     &+\beta'\left[\vec{\omega}_s\times\left(\frac{\rho_{nn}}{\rho_{s}}\vec{v}_s+\frac{\rho_{np}}{\rho_{s}}\vec{v}_p-\vec{v}_e+\frac{1}{\rho_s}\rot\vec{\lambda}_n\right)\right]+\\   \nonumber 
     &+\beta\vec{e}_k\times\left[\vec{\omega}_s\times\left(\frac{\rho_{nn}}{\rho_{s}}\vec{v}_s+\frac{\rho_{np}}{\rho_{s}}\vec{v}_p-\vec{v}_e+\frac{1}{\rho_s}\rot\vec{\lambda}_n\right)\right].
  \end{align}
  Here, $\Gamma_n$ is the superfluid neutrons creation rate, $\vec{\lambda}_n = \partial U_0/\partial\vec{\omega}_s$, $U_0$ is the neutron liquid internal energy density, $\vec{v}_e$ is the electrons velocity. It should be noted that the phenomenological approach by itself doesn't allow to estimate the mutual friction coefficients. Here, $\beta$ and $\beta'$ are just the arbitrary parameters.

  Mendell and Lindblom used the Khalatnikov's phenomenological approach but the same equations can be obtained from the vortices dynamics consideration if we suppose that the Magnus force takes the form
  \begin{equation}
    \vec{F}_m = \varkappa \vec{e}_k \times\left(\rho_s\vec{v}_l-\vec{J}_s-\rot\vec{\lambda}_n\right),
  \end{equation}
  and that the friction force equals to 
  \begin{equation}
    \label{eq:Mendell_fric_force}
    \vec{F}_c = -\frac{\rho_c}{n_v\tau_c}(\vec{v}_l-\vec{v}_e).
  \end{equation}
  The last assumption means that only the electrons scattering is taken into account. In present paper we restrict ourselves to considering only electron-scattering processes but in final equations the particular scattering mechanism specified only by $\beta$ and $\beta'$ coefficients. 
 
  As it follows from the calculations of \cite{Mendell1991a},
  \begin{equation}
    \label{eq:lambda_approximation}
    \vec{\lambda}_n \approx \frac{\varkappa}{4\pi}\frac{\varrho^2}{\rho_{pp}} \ln\left(\frac{l_v}{\xi_n}\right)\vec{e}_k,
  \end{equation}
  where $\varrho^2 = \rho_{nn}\rho_{pp}-\rho_{np}^2$ is the determinant of the mass density matrix.

  \section{The Model}\label{sec:model}
  We treat a neutron star core as a two-component system. The first component consists of the superfluid neutrons and its dynamics is described by equations \eqref{eq:Mendell_Euler}-\eqref{eq:Mendell_force}. Protons, electrons, and normal neutrons are coupled to each other on the small time-scales and form the second component which we denote as a ``charged'' component. All particles forming the charged component are supposed to move with the same velocity $\vec{v}_c$:
  \begin{equation}
    \label{eq:vc}
    \vec{v}_e = \vec{v}_{ex} = \vec{J}_{p(s)}/\rho_{p(s)} = \vec{v}_c.
  \end{equation}

  The star crust rotates with the angular velocity $\vec{\Omega}$ which is identified with the observed angular velocity of the pulsar. The external torque $\vec{K}$ acts on the crust. We will consider the torques depending only on $\vec{\Omega}$ and the configuration of magnetic field which supposed to be frozen into the crust. In this case, the torque $\vec{K}$ is a very slowly evolving vector in the frame co-rotating with the crust, so this frame is the most suitable for the considering problem. 

  First of all let us introduce the velocity fields and the mass currents measured in co-rotating frame:
  \begin{align}
    &\vec{u}_\alpha=\vec{v}_\alpha-[\vec{\Omega}\times\vec{r}], \\
    &\vec{J}_\alpha^* = \vec{J}_\alpha - \rho_\alpha[\vec{\Omega}\times\vec{r}].
  \end{align}
  Here, $\alpha$ is the constituent index ($\alpha = e, s, p(s), ex$). Moreover, assumption \eqref{eq:vc} allows us to introduce this quantities for the whole charged component in a similar form:
  \begin{align}
    &\vec{u}_c=\vec{v}_c-[\vec{\Omega}\times\vec{r}], \\
    &\vec{J}_c^* = \vec{J}_c - \rho_c[\vec{\Omega}\times\vec{r}],
  \end{align}
  where
  \begin{equation}
    \rho_c = \sum\limits_{\alpha\neq s} \rho_\alpha.
  \end{equation}
  It is important to note that by the $\vec{u}_c$ we will denote the quantity $\vec{J}_c^*/\rho_c$ (as it follows from \eqref{eq:vc}), so $\vec{u}_c$ is not equal to $\vec{u}_p$. 

  Equations \eqref{eq:Mendell_Euler}-\eqref{eq:Mendell_force} in co-rotating frame take the form
  \begin{align}
    \label{eq:co-rotation_Euler}  
    \partial^{*}_t\vec{u}_s+2[\vec{\Omega}\times\vec{u}_s]+(\vec{u}_s\cdot\nabla)\vec{u}_s+\nabla\tilde{\mu}_1
    = -[\dot{\vec{\Omega}}\times\vec{r}]+ \\
    +\frac{\rho_{np}}{\rho_{s}} (\vec{u}_p-\vec{u}_s)\times(2\vec{\Omega}+\rot\vec{u}_s)+\vec{f}_v,
    \nonumber
  \end{align}
  \begin{equation} 
    \partial^*_t\rho_s + \Div\vec{J}_s^* = \Gamma_s,
    \label{eq:co-rotation_continuity}
  \end{equation}
  \begin{align}
    \label{eq:co-rotation_force}  
     &\vec{f}_v=(2\vec{\Omega}+\rot\vec{u}_s)\times \left\{ -\frac{1}{\rho_s}\rot\vec{\lambda}_n + \right.\\ \nonumber
     &+ \beta'\left(\frac{\rho_{nn}}{\rho_{s}}\vec{u}_s+\frac{\rho_{np}}{\rho_{s}}\vec{u}_p-\vec{u}_e+\frac{1}{\rho_s}\rot\vec{\lambda}_n\right) + \\ 
    &\left.+\beta \frac{2\vec{\Omega}+\rot\vec{u}_s}{|2\vec{\Omega}+\rot\vec{u}_s|}\times\left(\frac{\rho_{nn}}{\rho_{s}}\vec{u}_s+\frac{\rho_{np}}{\rho_{s}}\vec{u}_p-\vec{u}_e+\frac{1}{\rho_s}\rot\vec{\lambda}_n\right)\right\},
    \nonumber     
  \end{align}
  where $\partial^{*}_t$ denotes the time derivative taken in the co-rotating frame, $\tilde{\mu}_1 = -(1/2)[\vec{\Omega}\times\vec{r}]^2+\tilde{\mu}+\Phi_G$, $\dot{\vec{\Omega}} = d_t\vec{\Omega}$.
  
\begin{figure}
    \includegraphics[height=54mm]{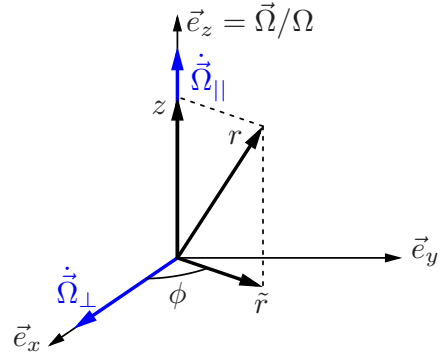}
    \caption{The cylindrical coordinate system and the vectors used in section \ref{sec:model}}
    \label{pic:dOmega}
 \end{figure}

  It will be useful further to expand $\dot{\vec{\Omega}} = \dot{\vec{\Omega}}_{||} + \dot{\vec{\Omega}}_\perp$, where $\dot{\vec{\Omega}}_{||} = (\dot{\vec{\Omega}}\cdot\vec{\Omega})\vec{\Omega}/\Omega^2$ and $\dot{\vec{\Omega}}_\perp = \dot{\vec{\Omega}} -\dot{\vec{\Omega}}_{||}$. 
  Based on this expansion let us introduce three orthogonal unit vectors $\vec{e}_x, \ \vec{e}_y, \ \vec{e}_z$ as follows:
  \begin{equation}
    \vec{e}_z = \frac{\vec{\Omega}}{\Omega}, \ \vec{e}_x = \frac{\dot{\vec{\Omega}}_\perp}{\dot{\Omega}_\perp}, \ \vec{e}_y = [\vec{e}_z \times \vec{e}_x].
  \end{equation}
  Let us also introduce associated cylindrical coordinates $\tilde{r},\phi,$ and $z$ ($\tilde{r}=0, \ z=0$ point the centre of the star) with corresponding orthogonal unit vectors $\vec{e}_{\tilde{r}}$, $\vec{e}_{\phi}$ and $\vec{e}_{z}$. We will use the notations: $\tilde{\vec{r}}$ for the cylindrical radius vector and  $\vec{r}$ for the spherical one (see fig. \ref{pic:dOmega}). All densities and coefficients are supposed to be functions only of $r$.
 
  We assume that the following inequalities are satisfied:
  \begin{equation}
    \label{eq:veloscity_inequality}
    \Omega \tilde{r} \gg u_s \gg u_c
  \end{equation}
  so we keep in equations \eqref{eq:co-rotation_Euler}-\eqref{eq:co-rotation_force} only the linear in $u_s$ terms and neglect all terms containing $u_e$ and $u_{ex}$. For the protons the particles (and charge) current is proportional to $\vec{J}_p$, so we suppose that 
  \begin{equation}
    \label{eq:Jp_asmpt}
    \vec{J}^*_{p(s)}/\rho_{p(s)}=\rho_e\vec{u}_{e}=\rho_{ex}\vec{u}_{ex} \approx 0.
  \end{equation}
  Together with \eqref{eq:sf_neutron_current} and \eqref{eq:sf_proton_current} it leads to
  \begin{equation}
    \label{eq:up}
    \vec{u}_p \approx - (\rho_{pn}/\rho_{pp})\vec{u}_s
  \end{equation}
  and
  \begin{equation}
    \label{eq:Js_asmpt}
    \vec{J}^*_s = \frac{\varrho^2}{\rho_{pp}}\vec{u}_s.
  \end{equation}
 
  The second part of inequality \eqref{eq:veloscity_inequality} requires a sufficiently effective physical mechanism damping the differential motions of the charged component and its strong coupling with the crust. It can be ensure by viscosity or by the magnetic field (the last requires the second type superconductivity for the protons). In any case, the more self-consistent model should involve hydrodynamical consideration for the both components of the star core but it significantly complicates the problem. We will include the $\vec{v}_c$ perturbation in our future developments.

  Let us discuss the $\rot\vec{\lambda}_n$-containing terms of equation \eqref{eq:Mendell_force} more detail. One can write 
  \begin{equation}
    \rot\vec{\lambda}_n = \frac{\partial\lambda_n}{\partial\tilde{r}}\vec{e}_\phi+\rot[\lambda_n(\vec{e}_k-\vec{e}_z)],
  \end{equation}
  where vector $\vec{e}_k-\vec{e}_z$ can be expanded in terms of $u_s$:
  \begin{equation}
    \vec{e}_k-\vec{e}_z\approx - \frac{1}{2\Omega}\vec{e}_z\times[\vec{e}_z\times\rot\vec{u}_s]+O\left(u_s^2\right).
  \end{equation}
  Using \eqref{eq:lambda_approximation}, one can obtain
  \begin{equation}
    \frac{1}{\rho_s}\omega_s\times\rot\vec{\lambda}_n\approx -\frac{2\Omega}{\rho_s}\frac{\partial\lambda_n}{\partial\tilde{r}}\vec{e}_{\tilde{r}}+O( u_s \epsilon)
  \end{equation}
  where we introduce a parameter
  \begin{equation}
    \epsilon\sim\frac{\varrho^2}{4\pi\rho_s\rho_{pp}}\left(\frac{\omega_s}{\Omega}\right)\left(\frac{l_v}{L}\right)^2\ln\left(\frac{l_v}{\xi_n}\right)
  \end{equation}
  which is much less than unity. Here, $L$ is the characteristic scale.
%   everywhere except maybe a thin layer near the superfluidity breaking surface
  In fact, it means that the linear in $u_s$ terms of the expansion of $\omega_s\times\rot\vec{\lambda}_n$ actually in addition to $u_s$ contains another small parameter $\epsilon$.
%   Linear in $u_s$ part of the expansion of $\rot\vec{\lambda}_n$-containing terms include the parameter $\epsilon\sim\lambda_n/\Omega\rho_s L^2$, where $L$ is the characteristic scale. Using \eqref{eq:lambda_approximation}, one can see that 
%   \begin{equation}
%     \epsilon\sim\frac{\varrho^2}{4\pi\rho_s\rho_{pp}}\left(\frac{\omega_s}{\Omega}\right)\left(\frac{l_v}{L}\right)^2\ln\left(\frac{l_v}{\xi_n}\right)\ll1
%   \end{equation}
%   everywhere except maybe a thin layer near the superfluidity breaking surface. 
  Hence, for the $\rot\vec{\lambda}_n$-containing terms we suppose that
  \begin{equation}
    \label{eq:rot_lambda_approx}
    \vec{\omega}_s\times\rot\vec{\lambda}_n\approx-\Omega\frac{\partial\lambda_n}{\partial\tilde{r}}\vec{e}_{\tilde{r}}
  \end{equation}
  is a good approximation. From a physical point of view it means that we neglect a part of vortices self-acting arising from their small curvature. 
  
  Assumption \eqref{eq:rot_lambda_approx} is a priori incorrect close to the superfluidity breaking surface. The vortices lines should satisfy the following boundary condition \citep{Khalatnikov_book}: 
%   for the velocity field to be finite:
   \begin{equation}
     \label{eq:perp_cond}
     [\vec{n}\times\vec{\omega}_s]_{r=r_s}=0,
   \end{equation}   
   where $r_{s}$ is the radius of sphere in which the superfluidity breaks (the radius of the superfluidity core) and vector $\vec{n}$ is normal vector to this surface. It means that $\vec{e}_k=\vec{n}$ on the surface and it significantly differs from $\vec{e}_z$ there. 
   However, the departure of $\vec{e}_k$ from $z$ direction exponentially decreases depthward the superfluid core with a characteristic scale 
%    However, the vortices direction deviation $\vec{e}_k-\vec{e}_z$ exponentially decreases depthward the superfluid core with characteristic scale   
  \begin{equation}   
     \ell \sim \sqrt{   
               \frac{1}{ \Omega}   
               \frac{\lambda_n}{\rho_{s}}   
             }   
     \sim 10^{-3} - 10^{-2} \mbox{cm}   
   \end{equation}
   It easy can be obtained using eq. (8) by \cite{SedrakyanSavvidi1979}
   So, this effect seems to be insignificant for our problem.

  Vector $\dot{\vec{\Omega}}$ as well as the external torque $\vec{K}$ evolves very slowly in co-rotating frame in compare with the mutual friction time-scales. Thus, the long-term evolution corresponds to the quasistationary solution of equations \eqref{eq:co-rotation_Euler} and \eqref{eq:co-rotation_continuity}. It means that the time-derivative terms $\partial^*_t$ become negligibly small on the large time-scales.

  Continuity equation \eqref{eq:co-rotation_continuity} can be rewritten as 
  \begin{equation}
    \partial^*_t\rho_s + \Div\vec{J}_s^* = \partial^*_t\rho_c + \Div\vec{J}^*_c.
  \end{equation}
  Here, $\partial^*_t\rho_s$ and $\partial^*_t\rho_c$ equal to zero as we just argued. Charged component mass current is negligibly small because of \eqref{eq:Jp_asmpt}. Thus, $\Gamma_s$ equals to zero up to corrections $\sim u_c$.   
  
  In present paper we also neglect the gravitational potential perturbation.

   After all simplifications we obtain the following equations for $u_s$:
  \begin{align}
     \label{eq:Euler}  
    2\Omega\beta\vec{e}_z\times\left[\vec{e}_z\times\left(\frac{\varrho^2}{\rho_{pp}\rho_s}\vec{u}_s-\frac{1}{\rho_s}\frac{1}{\tilde{r}}\frac{\partial\lambda_n}{\partial\tilde{r}}\vec{e}_\phi\right)\right] - \nabla\tilde{\mu}_1 - \\
   -2\Omega\gamma\left[\vec{e}_z\times\left(\frac{\varrho^2}{\rho_{pp}\rho_s}\vec{u}_s-\frac{1}{\rho_s}\frac{1}{\tilde{r}}\frac{\partial\lambda_n}{\partial\tilde{r}}\vec{e}_\phi\right)\right] = [\dot{\vec{\Omega}}\times\vec{r}] 
    \nonumber   
  \end{align}
  \begin{equation}
    \label{eq:continuity}
    \Div\left(\frac{\varrho^2}{\rho_{pp}}\vec{u}_s\right) = 0,
  \end{equation}
  where $\gamma = (1-\beta')$.
  In addition, the velocity field must satisfy the boundary condition
  \begin{equation}
    \label{eq:boundary_cond}
    [\vec{J}_n\cdot\vec{r}]_{r=r_{s}}\approx \left[\frac{\varrho^2}{\rho_{pp}}(\vec{u}_s\cdot\vec{r})\right]_{r=r_{s}}=0,
  \end{equation}
  Note that all mass densities in \eqref{eq:Euler}-\eqref{eq:boundary_cond} are not perturbed.

  The solution satisfying \eqref{eq:Euler}, \eqref{eq:continuity}, and \eqref{eq:boundary_cond} has the form
  \begin{align}
    \label{eq:Us_sol}
    \vec{u}_s = &\frac{\rho_{pp}\rho_s}{\varrho^2}[\vec{\varpi}\times\vec{r}]-\frac{\rho_{pp}\rho_s}{\varrho^2}\frac{\dot{\Omega}_{||}}{2\Omega}\frac{\gamma-\beta\psi}{\gamma^2+\beta^2}\vec{\tilde{r}}+ \\  \nonumber    
    &+\vec{e}_z\frac{\rho_{pp}}{\varrho^2}\frac{\dot{\Omega}_{||}}{2\Omega}\int\limits_0^z\frac{1}{\tilde{r}}\frac{\partial}{\partial\tilde{r}}\left(\tilde{r}^2\rho_s\frac{\gamma-\beta\psi}{\gamma^2+\beta^2}\right)dz',\\
    \label{eq:varpi_sol}
    \vec{\varpi} = &-\frac{\dot{\vec{\Omega}}_{||}}{2\Omega}\frac{\beta+\gamma\psi}{\gamma^2+\beta^2} - \frac{\beta}{\gamma^2+\beta^2}\frac{\dot{\vec{\Omega}}_\perp}{\Omega}+ \\     \nonumber
    &+\frac{\gamma}{\gamma^2+\beta^2}\left[\vec{e}_z\times\frac{\dot{\vec{\Omega}}_\perp}{\Omega}\right] +\frac{1}{\rho_s}\frac{1}{\tilde{r}}\frac{\partial\lambda_n}{\partial\tilde{r}}\vec{e}_z, \\ 
    \tilde{\mu}_1 = &-y z \dot{\Omega}_\perp - \dot{\Omega}_{||} \int\limits_0^{\tilde{r}} \psi(\tilde{r}')\tilde{r}'d\tilde{r}',\\
    \psi(\tilde{r}) = &\left[\int\limits_0^{z_b}\frac{\rho_s\gamma}{\gamma^2+\beta^2}dz'\right]\left[\int\limits_0^{z_b}\frac{\rho_s\beta}{\gamma^2+\beta^2}dz'\right]^{-1},\\
    z_b = &{\sqrt{r^2_{s}-\tilde{r}^2}}.
    \label{eq:psi_sol}
  \end{align}
  
  First, let us consider an uniform rotating star ($\dot{\vec{\Omega}} = 0$). In this case, the velocity field reduces just to 
  \begin{equation}
    \label{eq:us_nobraking}
    \vec{u}_s = \frac{\rho_{pp}}{\varrho^2}\frac{\partial\lambda_n}{\partial \tilde{r}}\vec{e}_\phi,
  \end{equation}
  but it is not equal to zero. It means that even in stationary situation the superfluid motion do not relaxes to rigid-body rotation. However, there is no paradox here. One just needs to recall that the vortices are the things which interact with the charged component but not the superfluid itself. It means that the friction tends to make $\vec{v}_l$ equal to $\vec{v}_e$. If we return to vortex force balance and suppose that $\vec{v}_l=\vec{v}_e$, we obtain
  \begin{align}
    \rho_s\varkappa\vec{e}_k\times\left(\vec{v}_l-\frac{\rho_{nn}}{\rho_{s}}\vec{v}_s-\frac{\rho_{np}}{\rho_{s}}\vec{v}_p-\frac{1}{\rho_s}\rot\vec{\lambda}_n\right) =\\ = -\frac{\rho_c}{n_v\tau_c}(\vec{v}_l-\vec{v}_e) = 0,
    \nonumber
  \end{align}
  or
  \begin{equation}
    \frac{\rho_{nn}}{\rho_{s}}\vec{v}_s+\frac{\rho_{np}}{\rho_{s}}\vec{v}_p-\vec{v}_e = -\frac{1}{\rho_s}\rot\vec{\lambda}_n+V_{||}\vec{e}_k,
  \end{equation}
  where $V_{||}$ is an arbitrary function. Taking into account equations \eqref{eq:up} and \eqref{eq:rot_lambda_approx}, we obtain exactly equation \eqref{eq:us_nobraking} ($V_{||}$ equals to zero because of continuity equation \eqref{eq:continuity} and boundary condition \eqref{eq:boundary_cond}). 
  So, the mass density gradient in star core gives rise to the force pushing vortices from the rotational axis.
%   The gradient of mass density in star core leads to force pushing vortices from the rotational axis.
  
  In general case, the flow has a complex form. It consists of a differential rotation about the local axis $\vec{\varpi}$ and a poloidal flow. The last, however, is the consequence of the variation of the mutual friction coefficients with $r$ and it vanishes if we make $\beta$ and $\beta'$ to be constant. If we, in addition, neglect the entrainment effect ($\rho_{pp}\rho_s/\varrho^2\approx 1$) and the vortices energy ($\lambda_n\approx 0$), we obtain a rigidly rotating solution. This simple case has been studied by the authors in their previous paper \citep{BGT2013}.
  
  The inequality $u_s\ll\Omega r$ must be satisfied for the solution to be valid. If $\beta$ and $\beta'$ can be expressed in terms of coupling parameter $\sigma$ according to \eqref{eq:betas_def}, taking into account that in this case
  \begin{equation}
     \frac{\gamma}{\gamma^2+\beta^2} = 1, \ \ \frac{\beta}{\gamma^2+\beta^2} = \sigma,
  \end{equation}  
  we obtain the following conditions for $\sigma$:
  \begin{equation}
     \dot{\Omega}_{||}/\Omega^2\ll\sigma\ll \Omega^2 / \dot{\Omega}_\perp.
  \end{equation}
  On the one hand, the vortices lattice should interact with charged component efficiently enough to rapidly reform with respect to the changing of $\vec{\Omega}$. One the other hand, if the interaction will be very strong, the vortices can not move through the charged component almost at all. The absence of the interaction as well as the very strong coupling leads to that the difference between $\vec{v}_s$ and  $\vec{v}_c$ during the rotation evolution can become not small.
  
  Protons of course are not required to be superconducting  everywhere inside the superfluid core. Entrainment effect vanishes where the superconductivity breaks \citep{GusakovHaensel2005}. Obtained solution, however, remains valid there. One just needs to replace
  \begin{equation}
    \label{eq:densities_replacements}
    \frac{\rho_{pp}\rho_s}{\varrho^2} \rightarrow 1\ \ \mbox{and} \ \ \frac{\rho_{pp}}{\varrho^2} \rightarrow \frac{1}{\rho_s}.
  \end{equation}
  Except this formally replacements the superconductivity breaking apparently leads to the mutual friction coefficients decrease by several orders of magnitude (see the detail below). 

  The solution formally tends to infinity on the superfluid breaking surface. It means that $u_s$ can not be treated as a small perturbation in the boundary region and one needs to solve the non-linear hydrodynamical equations there. However, the thickness of the layer in which the non-linear terms become important seems to be very small. Moreover, the superfluid mass current which we will use in next section to formulate the equations of pulsar motion obtained with \eqref{eq:Us_sol}-\eqref{eq:psi_sol} tends to zero there. So, this layer seemingly does not play a significant role in the angular momentum transfer.

  \section{The equations of motion} \label{sec:eq_of_motion}
  Possessing the expression for the superfluid velocity field, one can calculate the angular momentum transfer rate from superfluid core to the charged component. From the angular momentum conservation law we have
   \begin{equation}
     \label{eq:Mc_eq}
     \dot{\vec{M}} =  \vec{K},
  \end{equation} 
  where $\vec{M}$ is the full angular momentum of the star, which obviously is just a sum of angular momenta of the components. So one can write
  \begin{equation}
    \vec{M} = \vec{M}_c + \vec{M}_s = \int_{ns}\vec{r}\times(\vec{p}_c+\vec{p}_s)dV,
  \end{equation}
  and, using \eqref{eq:J&p_eq},
  \begin{equation}
    \vec{M} = \int_{ns}\vec{r}\times(\vec{J}_c+\vec{J}_s)dV.
  \end{equation}
  The taking into account \eqref{eq:Jp_asmpt} and \eqref{eq:Js_asmpt} after some integrating leads to 
  \begin{equation}
    \label{eq:dOm}
    I_c\dot{\vec{\Omega}} = - I_s\dot{\vec{\Omega}} -\int_{sf}\frac{\varrho^2}{\rho_{pp}}\left[\vec{r}\times\frac{\partial\vec{u}_s}{\partial t}\right]dV+\vec{K},
  \end{equation}
  where we introduce the moments of inertia for the components
  \begin{equation}
    I_c = \frac{8\pi}{3}\int\limits_0^{r_{ns}}\rho_c r^4 dr + I_{cr}, \ \     I_s = \frac{8\pi}{3}\int\limits_0^{r_{ns}}\rho_s r^4 dr,
  \end{equation}
  where $\rho_s$ and $\rho_c$ are the mass densities of the superfluid and charged components, $I_{cr}$ is the crust moment of inertia.

  The next step is to differentiate the velocity field $\vec{u}_s$ with respect to time. Doing this, we as before neglect $\partial^*_t \vec{u}_s$ term, so 
  \begin{equation}
    \label{eq:dus}
    \partial_t \vec{u}_s \approx [\vec{\Omega}\times\vec{u}_s] - \left([\vec{\Omega}\times\vec{r}]\cdot\nabla\right)\vec{u}_s.
  \end{equation}
  Substituting \eqref{eq:Us_sol},\eqref{eq:varpi_sol} into \eqref{eq:dus}, and \eqref{eq:dus} into \eqref{eq:dOm}, we obtain
  \begin{equation}
    \label{eq:dOm_eq}
    I_c\dot{\vec{\Omega}} = S_1 I_s \dot{\vec{\Omega}} -S_2 I_s\dot{\vec{\Omega}}_{||} + S_3 I_s[\vec{e}_z\times\dot{\vec{\Omega}}] +\vec{K},
  \end{equation}
  where we introduce the following coefficients:
  \begin{align}
    &S_1 = \frac{8\pi}{3I_s}\int\limits_0^{r_{s}}\frac{\gamma\beta' - \beta^2}{\gamma^2+\beta^2}\rho_s r^4 dr, \\ &S_2 = \frac{8\pi}{3I_s}\int\limits_0^{r_{s}}\frac{\gamma}{\gamma^2+\beta^2}\rho_s r^4 dr, \\ 
    &S_3 = \frac{8\pi}{3I_s}\int\limits_0^{r_{s}}\frac{\beta}{\gamma^2+\beta^2}\rho_s r^4 dr, 
  \end{align}
  Note that if expressions \eqref{eq:betas_def} are correct, $S_1$ equals to zero and $S_2$ equals to unity. Equation $S_2-S_1 = 1$, however, is satisfied without any assumptions about the relations between $\beta$ and $\beta'$.

\begin{figure}
    \center\includegraphics[height=54mm]{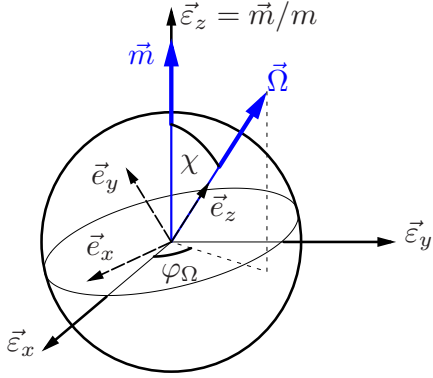}
    \caption{The vectors used in section \ref{sec:eq_of_motion}}
    \label{pic:vectors}
  \end{figure}

  Next, we need to introduce a new basis $\vec{\varepsilon}_x$, $\vec{\varepsilon}_y$, and $\vec{\varepsilon}_z$, where $\vec{\varepsilon}_z = \vec{m}/m$ and vectors $\vec{\varepsilon}_x$ and $\vec{\varepsilon}_y$ are anchored in the star crust and perpendicular to $\vec{\varepsilon}_z$ and to each other. In this basis the orientation of $\vec{\Omega}$ determines by two angles $\chi$ and $\varphi_\Omega$ (see fig. \ref{pic:vectors}). Angle $\chi$ is the pulsar inclination angle, and the variation of $\varphi_\Omega$ relates with star precession. Without making any additional restriction external torque $\vec{K}$ can be represented as
  \begin{equation}
    \vec{K} = K_\Omega \vec{e}_z+K_m \vec{\varepsilon}_z+K_\perp[\vec{e}_z\times\vec{\varepsilon}_z]
  \end{equation} 
  Equation \eqref{eq:dOm_eq} can be solved for $\dot{\vec{\Omega}}$ and rewritten as three scalar equations which describe pulsar braking, inclination angle evolution and torque-driven precession respectively:
  \begin{equation}
    \label{eq:omega_eq}
    \dot{\Omega} = \frac{K_\Omega+K_m\cos\chi}{I_s+I_c},
  \end{equation}
  \begin{equation}
    \label{eq:chi_eq}
    \dot{\chi} = -\frac{1}{\Omega}\frac{(I_c-S_1I_s)K_m -S_3I_sK_\perp}{(I_c-S_1I_s)^2+S_3^2I_s^2}\sin\chi,
    \end{equation}
  \begin{equation}
    \label{eq:phi_eq}
    \dot{\varphi}_\Omega = -\frac{1}{\Omega}\frac{(I_c-S_1I_s)K_\perp+S_3I_sK_m}{(I_c-S_1I_s)^2+S_3^2I_s^2}.
  \end{equation}
  Equation \eqref{eq:omega_eq} formally has the same form as the one for the rigid star \citep{BarsukovPolyakovaTsygan2009}. Neglecting $\partial^*u_s$, we assume, in fact, that the energy loss of the superfluid component equals just to $(I_s/2)(\vec{\Omega}\cdot\dot{\vec{\Omega}})$.
  However, it doesn't mean that the breaking will occur with the same rate. Equations \eqref{eq:omega_eq}-\eqref{eq:phi_eq} should be solved simultaneously. The presence of the superfluid core leads to that the torque perpendicular term $K_\perp$ starts to affect the inclination angle evolution (see \cite{CasiniMontemayor1998}) as well as term $K_m$ starts to affect the precession.

  \section{Application}\label{sec:application}
  Up to now the developed formalism does not require any specification of the profiles of the mutual friction coefficients and the mass densities as well as $K_\Omega$, $K_m$ and $K_\perp$ can be arbitrary small in magnitude slow varying functions. Next we apply the particular models (simplistic at some points) in order to demonstrate that the choice of the model may significantly affect the rate of the neutron star rotation evolution.
  
  Mass density matrix $\rho_{\alpha\beta}$ with the taking into account the temperature effects has been calculated by \citet{GusakovHaensel2005}. We do not give here the lengthy expressions obtained by Gusakov and Haensel.
  Note, however, that the densities $\rho_{\alpha\beta}$ depend on the parameters $\tau_\alpha = T/T_{c\alpha}$, where $T$ is the mixture temperature and $T_{c\alpha}$ are the critical temperatures of proton and neutron fluids. Gusakov and Haensel have shown that the magnitudes of $\rho_{\alpha\alpha}$ and $\rho_{\alpha\beta}$ tend to zero with $T$ approaching $T_{c\alpha}$. It means that the entrainment effect vanishes when the superfluidity breaks for one of the fluids. We have used this in \eqref{eq:densities_replacements}.
  
  The critical temperatures themselves depend on the mass density. Unfortunately, these dependences calculated by different authors may significantly differ from each other \citep{YakovlevLevenfishShibanov1999}. It may leads to the large variation of the amount of the neutron superfluid in star cores and the mutual friction coefficients. The aim of this section is to investigate the sensitivity of the long-term rotation dynamics to these variations.
    
  Despite the differences between the nucleon superfluidity/superconductivity theories there are some common results \citep{YakovlevLevenfishShibanov1999}.
%   In any case, 
  Each critical temperature profile should possess a maximum $T_{c\alpha}^{max}$ at some density $\rho_{c\alpha}^{max}$. For the protons  $\rho_{cr}^{max}$ exceeds the nuclear density $\rho_0\approx2.8\times10^{14}$~g/cm$^3$. For the neutrons, actually, there are two types of the neutron superfluidity. Singlet-pairing  ($^1S_0$) takes place in the neutron stars crusts. We neglect this phenomenon in present paper. In the star core the microscopic theories predict a triplet-pairing type ($^3 P_2$) superfluidity for which $\rho_{cn}^{max}>\rho_{cp}^{max}$ and $T_{cn}^{max}<T_{cp}^{max}$.

\begin{figure}
    \center\includegraphics[height=48mm]{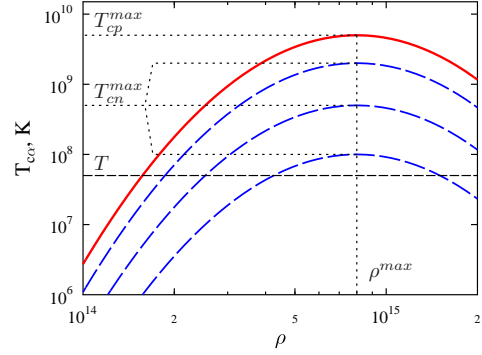}
    \caption{The critical temperature profiles for protons (thick solid line) and neutrons (thick dashed lines).}
    \label{pic:Tcr_profs}
  \end{figure}

  We model the critical temperature profiles by the parabolas (see fig \ref{pic:Tcr_profs}):
  \begin{equation}
    \label{eq:Tc_profs}
    \log_{10}\left(\frac{T_{c\alpha}}{T^{max}_{c\alpha}}\right) = -4\log^2_{10}\left(\frac{\rho}{\rho^{max}_{\alpha}}\right).
  \end{equation} 
  We fix $T_{cp}^{max}$ to be equal to $5\times10^9$ K and variate $T_{cn}^{max}$. The densities $\rho^{max}_{p}$ and $\rho^{max}_{n}$  corresponding to the maximum of the profiles both are put to be equal to $8\times10^{14}$g/cm$^3$.
  It is not quite realistic of course but the variation of $T_{cn}^{max}$ allows us to investigate the sensitivity of the results to the choice of the critical temperature profile.
  
  We will consider a neutron star with radius $r_{ns}\approx12$ km. Parameter $r_{s}$ is determined as a radius of the sphere on which $T = T_{cr(n)}$. The star core is assumed to be isothermal. Its temperature does not change with time and equals to $5\times10^7$ K. In order to calculate $\rho_p$ and $\rho_n$ profiles we use \citet{HeiselbergHjorth-Jensen1999} approximation of ARP98 equation of state.
  
  According to \eqref{eq:proton_current} the neutron vortices must generate the magnetic field with typical fluxes which equal to
  \begin{equation}
    \Phi_* = \left(\frac{m_p}{m_n}\right)\left(\frac{\rho_{np}}{\rho_{pp}}\right)\Phi_0,
  \end{equation}
  where $\Phi_0 = 2\pi\hbar c/2e$ is the flux quantum \citep{AlparLangerSauls1984}. We suppose that the most effective mutual friction mechanism is based on the electrons scattering on the vortices magnetic field. For this mechanism the coupling parameter has been calculated by \citet{AlparLangerSauls1984} and it equals to
  \begin{equation}
    \label{eq:alpar_sigma}
    \sigma_{ALS} \approx 1.3\times10^{-2}\frac{x}{1-x} \left(x\rho_{14}\right)^{1/6}\left(\frac{m_p}{m_p^*}\right)^{1/2}\left(\frac{\rho_{np}}{\rho_{pp}}\right)^2,   
  \end{equation}
  where $x = \rho_c/\rho$, $\rho_{14} = \rho/10^{14}$g/cm$^{3}$, $\rho=\rho_c+\rho_s$, $m_p^*$ is the effective proton mass. Thus, it is the weak-coupling regime, so \begin{equation}
    \label{eq:coupl_coeffs}
    S_1=0, \ \ S_2=1, \ \ S_3\ll1.                                                                                                                                                                                                                                                               
  \end{equation}
  Not that this mechanism requires the protons to be in superconducting state. In our simplistic model it satisfies everywhere inside the superfluid neutron core. 
%   If it is not so in the real neutron star, 
  The neutron vortices have also the own magnetic field caused by the spontaneous magnetization which should take place in $^3P_2$ neutron superfluid liquids \citep{SaulsSteinSerene1982}. However, this field is several orders of magnitude smaller than the field generated by proton currents. So, if the superconductivity breaks in some core region, the mutual friction become much less effectively.
  The electrons also can scatter on the thermal excitation in the vortices cores \citep{Feibelman1972} This mechanism is very sensitive to the temperature, however, it apparently is much less effective then the Alpar-Lauger-Sauls scattering even for the young neutron stars. Except the electrons the neutron vortices can interact with the proton flux tubes. However, we will suppose either that the superconducting protons are in the first type state or that this kind of interaction is much less effective then the Alpar-Lauger-Sauls scattering too.

  As for the torque $\vec{K}$, we use the one proposed by \citet{BarsukovPolyakovaTsygan2009}:
  \begin{align}
    \label{eq:BPT_torque}
    \vec{K} = K_0\big(\left(1-\alpha(\chi,\varphi_\Omega)\right)\cos\chi\vec{e}_m - \vec{e}_z + \\ +R\cos\chi[\vec{e}_z\times\vec{e}_m] \big)
    \nonumber
  \end{align}
  where
  \begin{equation}
      K_0 = \frac{2\Omega^3m^2}{3c^3}, \ \ R = \frac{9}{10}\frac{c}{\Omega r_{ns}}. 
  \end{equation}
  This torque is a sum of magneto-dipolar angular momentum losses torque calculated in the vacuum approximation \citep{Deutsch1955} and the one related with the currents losses \citep{Jones1976}. Here, $\alpha(\chi,\varphi_\Omega)$ is the function $\sim1$ which depends on the structure of the small-scale magnetic fields in vicinities of neutron star magnetic poles. Small-scale fields bend the pulsar tubes and affect the tubes currents. In the case of pure dipolar magnetic field, function $\alpha(\chi,\varphi_\Omega)$ is a constant. 
%   [and equals to $1.275$].
  
  The last term in \eqref{eq:BPT_torque} leads to the torque-driven radiative precession \citep{Melatos2000}. Note that this term contains the big parameter
  \begin{equation}
    R\approx3.6\times10^3\left(\frac{P}{1\mbox{ sec}}\right),
  \end{equation}
  so $K_\perp\gg K_m$ and $K_\Omega$ for this torque. Together with \eqref{eq:coupl_coeffs} it means that the first term in \eqref{eq:phi_eq} much greater than the second and, consequently,
  \begin{equation}
    \dot{\varphi}_\Omega \approx -\frac{I_cK_0R}{I_c^2+S_3^2I_s^2}\cos\chi.
%     \left(\frac{9c}{10r_{ns}\Omega}\right)
  \end{equation}

\begin{figure}
    \center\includegraphics[height=50mm, trim= 0.9cm 0.8cm 1.7cm 1.4cm, clip = true]{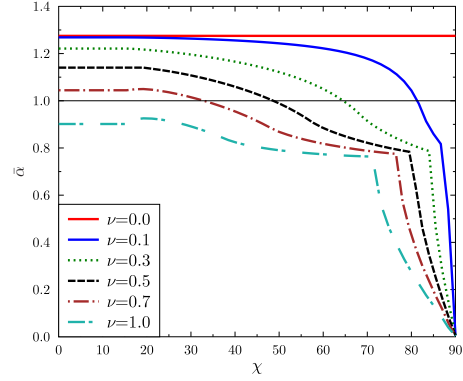}
    \caption{The behaviour of function $\bar{\alpha}_\nu(\chi)$ for different values of non-dipolarity parameter $\nu$.}
    \label{pic:alpha}    
  \end{figure}

  If $\chi$ is not too close to $\pi/2$, the radiative precession period 
  \begin{align}
    T_p &= \frac{2\pi}{K_0R}\frac{I_c^2+S_3^2I_s^2}{I_c\cos\chi} \approx \\ &\approx 1.9\times10^{5}\left(\frac{P}{1\mbox{ sec}}\right)\frac{I_{c(45)}^2+S_3^2I_{s(45)}^2}{B_{12}I_{c(45)}\cos\chi} \mbox{ years}
    \nonumber
  \end{align}
   is three orders of magnitude smaller that the characteristic time-scale of the pulsar braking and inclination angle evolution. Here, $I_{\alpha(45)}$ are the moments of inertia measured in units of $10^{45}$ g/cm$^3$, $B_{12}$ is the dipolar magnetic field measured on the magnetic poles in units of $10^{12}$ G.  This fact allows us to average the equations over $T_p$. Dividing averaged equation \eqref{eq:chi_eq} by averaged \eqref{eq:omega_eq}, taking into account that $P = 2\pi/\Omega$, one can obtain (see \citet{BGT2013} for the detail)
  \begin{align}
    \nonumber
    \frac{d\chi}{dP} \approx -\frac{1}{P}\frac{I_c(I_c+I_s)}{I_c^2+S_3^2I_s^2}\frac{\sin\chi\cos\chi}{\sin^2\chi+\bar{\alpha}_\nu(\chi)\cos^2\chi}\times \\
     \times\left[1-\bar{\alpha}_\nu(\chi)-S_3\frac{I_s}{I_c} \frac{9}{20\pi} \left(\frac{c}{r_{ns}}\right)P\right],
     \label{eq:dChi_dP}
  \end{align} 
  where $\bar{\alpha}_\nu(\chi) = (2\pi)^{-1}\int_0^{2\pi}\alpha(\chi,\varphi_\Omega)d\varphi_\Omega$.
%    The behaviour of function $\bar{\alpha}_\nu(\chi)\mathbf{ = (2\pi)^{-1}\int_0^{2\pi}\alpha(\chi,\varphi_\Omega)d\varphi_\Omega}$ depends on the non-dipolarity parameter $\nu=B_1/B_0$, where $B_0$ and $B_1$ are the magnitudes of dipolar and small-scale magnetic field measured on the neutron star polar caps. 
%    \textbf{Not any bended pulsar tube orientation relative to the rotation axis is compatible with the existence of the stable tube current. So, it leads to decrease of the current value averaged over $T_p$ and, consequently, to decrease of $\alpha(\chi,\varphi_\Omega)$ averaged over $\varphi_\Omega$.}  
   The behaviour of $\bar{\alpha}(\chi)$ calculated in the framework of the model proposed by \citet{BGT2013}  is shown in figure \ref{pic:alpha}. In proposed model the behaviour of $\bar{\alpha}_\nu(\chi)$ is determined only by non-dipolarity parameter $\nu=B_1/B_0$, where $B_0$ and $B_1$ are the magnitudes of dipolar and small-scale magnetic field measured on the neutron star polar caps.
%    for different $\nu$. 
  
   Equation $\bar{\alpha}(\chi)=1$ has a root $\chi_{eq}$ for a range of non-dipolarity parameter $\nu$ values. For a perfectly rigid star $\chi_{eq}$ is the equilibrium inclination angle. 
   
   Inclination angle evolution as a function of the rotation period does not depend on the magnitude of the magnetic field but only on the relation $\nu = B_1/B_0$. 
%    Thus, there is no need to include in the consideration the magnetic field decay model
   We suppose that non-dipolarity parameter $\nu$ remains constant during the evolution. This assumption supported by some magnetic decay models according to which the parameter $\nu$ evolves much slower than the magnetic field itself \citep{MitraKonarBhattacharya1999} (We are interested in the small-scale fields with characteristic scale $\sim 1$ km). However, it should be noted that it is not the generally accepted result \citep{UrpinGil2004}
  
  The last term of equation \eqref{eq:dChi_dP} (see also eq. \eqref{eq:chi_eq}) contains $K_\perp$ which is not affect the inclination angle evolution in the case of rigidly rotating star.
%   \textbf{The presents of the superfluid core results in that} $K_\perp$ starts to affect the inclination angle evolution. 
  The influence of this term becomes important when the rotation period approaches to the value
  \begin{equation}
    P\sim \frac{I_c}{S_3 I_s}\frac{20\pi r_{ns}}{9 c}.
  \end{equation}
  So, the greater the amount of the neutron superfluid and the stronger the interaction, the sooner it happens. When this term becomes dominant, it makes all pulsars evolve to the orthogonal state.
  
\begin{figure} \center
    \includegraphics[height=45mm]{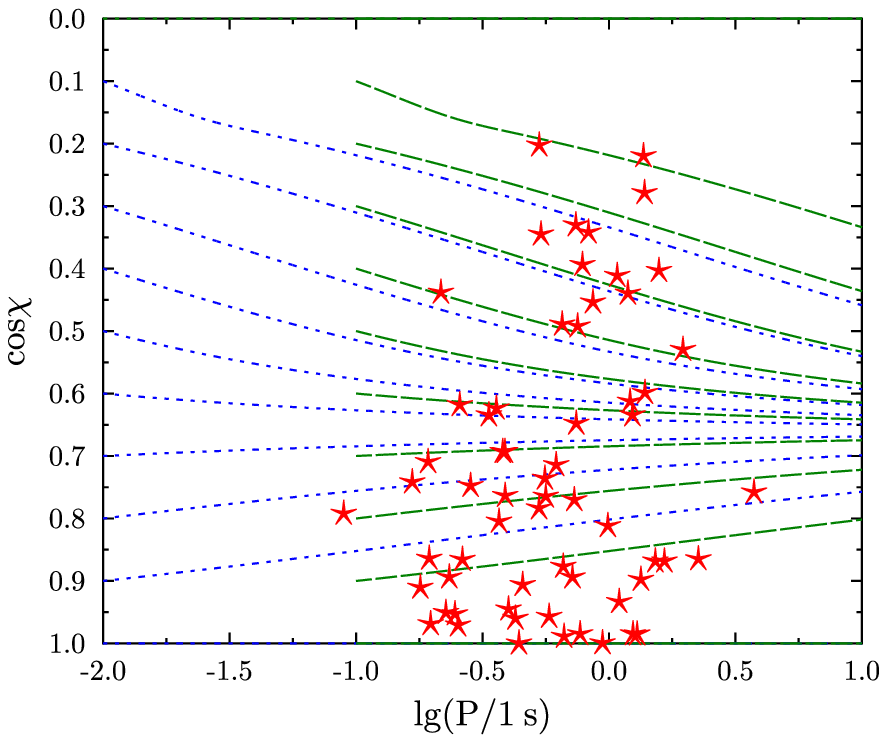} \\
     a \\ \ \\
    \includegraphics[height=45mm]{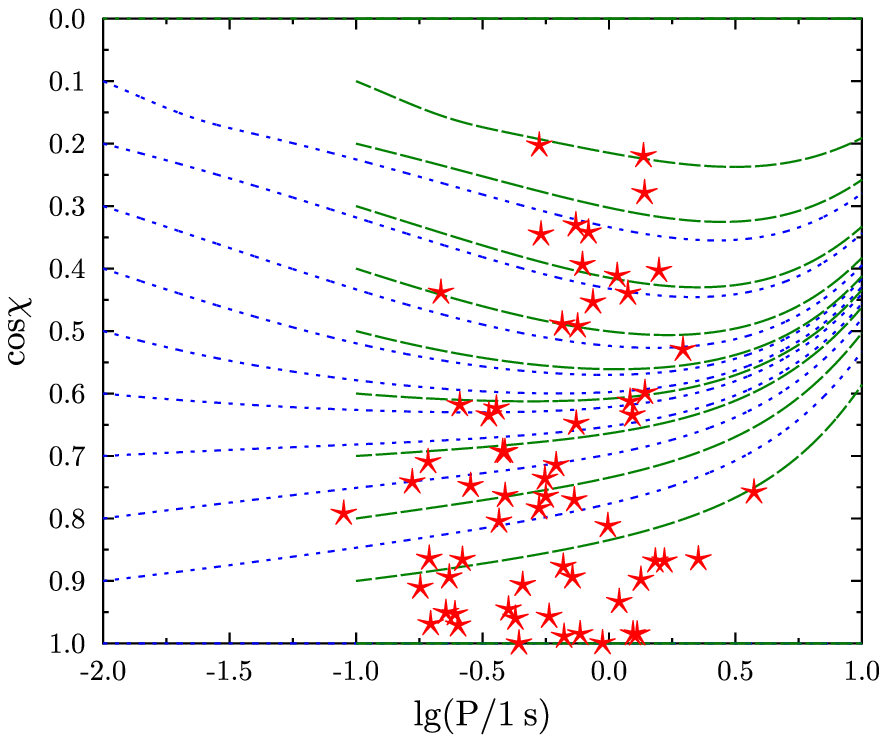} \\
    b \\ \ \\
    \includegraphics[height=45mm]{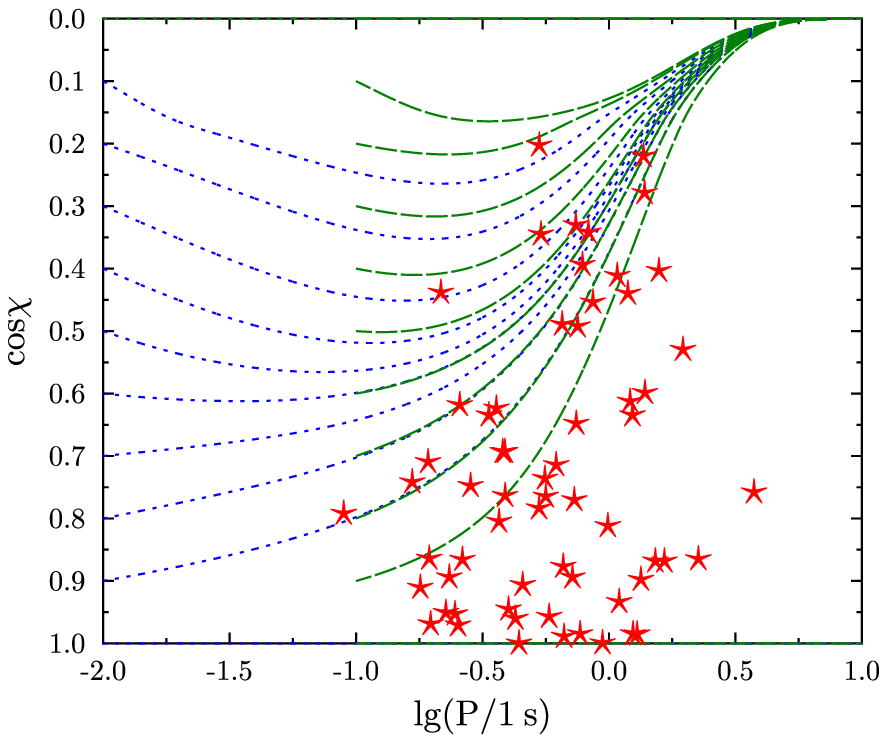}\\
    c\\ \ \\
    \includegraphics[height=45mm]{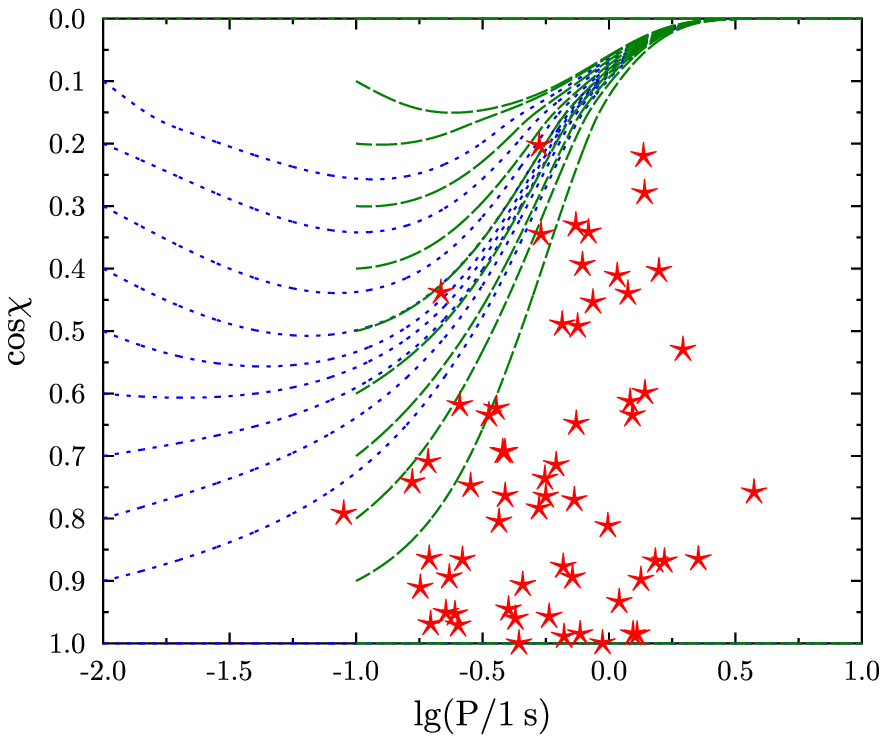} \\
    d\\
    \caption{
%     The evolution trajectories obtained for non-dipolarity parameter $\nu=0.5$ and $T^{max}_{cn} =1\times10^{8}$ K (top panel), $5\times10^{8}$ K (middle panel), $2\times10^{9}$ K (bottom panel) for pulsars with initial periods equal to 10 msec (dotted lines) and 100 msec (dashed lines). Stars demonstrate the observation data for 62 pulsars from \citet{Malov_book}
      The evolution trajectories obtained (a) for the rigidly rotating star ($I_s=0$) and for the ``two components'' star with $T^{max}_{cn}$ equal to (b) $1\times10^{8}$ K, (c) $5\times10^{8}$ K (d) $2\times10^{9}$ K. Dotted lines correspond to the initial period $P_0$ equal to 10 msec, and dashed lines correspond to $P_0=100$ msec. The non-dipolarity parameter $\nu$ is taken to be equal to $0.5$.
    }
    \label{pic:trajectories}
  \end{figure}

  The calculated values of $I_s/(I_s+I_c)$ and $S_3$ for different $T^{max}_{cn}$ are given in table \ref{tab:values}.
  The evolution trajectories for different initial $\chi$ and $P$ and different $T^{max}_n$ obtained by the integration of equation \eqref{eq:dChi_dP} are shown in figure \ref{pic:trajectories}. 
  
  The observational data on 62 pulsars also are plotted in the figure. We use the data from \citep{Malov_book} on pulsars inclination angles measured by using the maximum derivative of the position angle (denoted as $\beta_2$ is the book).
  \begin{table}
    \caption{The values of $I_s/(I_s+I_c)$ and $S_3$ for different $T^{max}_{cn}$} \center
    \begin{tabular}{ccc}
      \hline
      $T^{max}_{cn}$ ($\times10^8$K)& $I_s/(I_s+I_c)$ & $S_3\ (\times10^{-5})$ \\
      \hline
                   $1$ &  $0.07$ &$2.38$\\ 
                   $5$ & $0.33$ &$5.08$\\
                   $20$ &  $0.5$ &$5.05$\\
      \hline
    \end{tabular}
    \label{tab:values}
  \end{table}

  \section{Discussion}\label{sec:discussion}
  
  The semi-hydrodynamical consideration
%   leads to 
  results in equation \eqref{eq:dChi_dP} which is equivalent to the analogous equation obtained under the  assumption that the mutual friction coefficients are constant and both components rotate rigidly \citep{BGT2013}. One just need to replace $\sigma$ by its value averaged over the core volume with the weight function $\sim \rho_s r^2$. As one can see from table \ref{tab:values} the value of $S_3$ does not change significantly with $T^{max}_n$. However, the higher values of $T^{max}_n$ correspond to higher values of relation $I_s/(I_s+I_c)$ and to the more rapid inclination angle evolution. 
  
%   \textbf{The rigidly rotating stars with the small-scale magnetic fields have the equilibrium inclination angles for a range of non-dipolarity parameters \citep{BarsukovPolyakovaTsygan2009}}
%   Moreover,
  In more general case, a root of equation $\bar{K}_m(\chi_{eq})=0$ is the stable inclination angle when $\frac{d\bar{K}_m}{d\chi}(\chi_{eq})>0$  for the rigidly rotating star (the bar denotes the averaging over $\varphi_\Omega$). Even if the rate of the angle evolution is high, the observed wide distribution of $\chi$ can be explained by the existence of the equilibrium inclination angles which may be different for each pulsar. For example, $\nu$ should vary from one pulsar to another.
%   \textbf{Putting  $I_s$ to be equal to zero, from equation \eqref{eq:chi_eq} one can see that for the rigidly rotating stars the root of the equation $\bar{K}_m(\chi_{eq})=0$ is a stable equilibrium inclination angle when $d\bar{K}_m/d\chi>0$ (bar denotes the averaging over $\varphi_\Omega$). Even if the rate of the angle evolution is high, the observed wide distribution of $\chi$ can be explained by the existence of the equilibrium inclination angles}
%   The rigidly rotating stars may possess a stable equilibrium inclination angle $\chi_{eq}$
%   \textbf{Even} if the rate of the evolution is high \textbf{(due to some reasons)}, the observed \textbf{wide} distribution of $\chi$ can be explained for the rigidly rotating  stars by the existence of the equilibrium inclination angles caused by, for example, the small-scale magnetic fields  \citep{BarsukovPolyakovaTsygan2009}. 
  But now, as it follows from equation \eqref{eq:dChi_dP}, the superfluid core makes the pulsars evolve to the orthogonal state and its influence increases during the star breaking. So, fast evolution, corresponding to the large amount of the superfluid neutrons, seems to be in contradiction with the observational data. This facts may allow to examine the superfluid models. 
  Note, however, that we've considered a perfect spherical star. The oblateness along the magnetic axis which should be caused by strong magnetic field \citep{Goldreich1970} just leads to the redefinition of the $R$ parameter \citep{BarsukovTsygan2010}. If the oblateness is strong enough, parameter $R$ may become negative, and the pulsars will tend to coaxial state instead the orthogonal. But it doesn't change the situation qualitatively. Triaxial stars, in principle, may possess the equilibrium inclination angles different from $0$ and $\pi/2$. This question requires a separate studying.
  
  In section \ref{sec:application} we've supposed that the protons are in the superconducting state. If the state is of the first type \citep{BuckleyMetlitskiZhitnitsky2004}, there is no magnetic field inside the superfluid core. In the case of the second type superconductivity \citep{BaymPethickPines1969a}, magnetic field contains in the flux tubes each of which carries the quantum of the magnetic flux $\Phi_0=hc/2e$. The flux tubes density can be estimated $n_f\sim B/\Phi_0 = 5\times 10^{19} B_{13}$ cm$^{-2}$, where $B_{13}$ is the averaged over the area $\gg n_f^{-2/3}$ magnetic field in units of $10^{13}$G. Meanwhile, averaging \eqref{eq:v_def} over the area $\gg l_v^2$, one can obtain $n_v\sim 6.4\times10^3 P^{-1}$cm$^{-2}$. So, flux tubes much more numerous than the neutron vortices. It allows to replace the flux tubes lattice by the continuum medium when a vortex dynamics is considered. If the core magnetic field has the chaotic structure \citep{RudermanZhuChen1998}, the interaction with the 
flux tubes can be included in 
the model just by replacing $\sigma_{ALS}$ by $\sigma_{ALS}+\sigma_{ft}$, where  
  $\sigma_{ft}$ describes the vortex friction on the flux tubes medium. 
  The calculations show \citep{SideryAlpar2009} that the interaction with flux tubes much more effective ($\sigma_{ft}\gg\sigma_{ALS}$), so the inclination angle evolution should be more sensitive to the relation $I_s/I_c$.
  If the magnetic field structure has a preferred direction (for example, if it has large toroidal component \citep{Braithwaite2009}), the friction becomes anisotropic. That case goes beyond the scope of the model proposed in present paper.

  We've assumed that $u_s\gg u_c$. In fact, it is a strong assumption, which is not quite established in present paper. The more self-consistent consideration requires the investigation of the influence of the charged component velocity perturbation on the rotational dynamics along with $u_s$.
  We do not take into account any pinning phenomena which may play significant role in neutron star rotational dynamics. Also we do not take into account the thermal evolution of the neutron stars. We are planing to include this factors in our future developments.

  \section*{Acknowledgements}  
  The authors are grateful to M.~E.~Gusakov and E.~M.~Kantor for helpful discussion. The authors are also would like to thank the referee for the useful comments. This work was supported by the Russian Foundation for the Basic Research (project 13-02-00112), the Program of the State Support for Leading Scientific Schools of the Russian Federation (grant NSh-4035.2012.2) and Ministry of Education and Science of Russian Federation (Agreement No.8409, 2012 and contract No. 11.G34.31.0001). 
  
  \bibliography{mn-jour,paper}   

\end{document}